\documentclass[prl,aps,twocolumn,showpacs,superscriptaddress,amsfonts,amsmath,floatfix]{revtex4}

\usepackage{graphicx}  
\usepackage{dcolumn}   
\usepackage{bm}      
\usepackage{amssymb}  
\usepackage{color}

\begin{document}

\title{Berry phase for a Bose gas on a one-dimensional ring}
\date{\today}

\author{Marija Todori\'{c}}
\thanks{mtodoric@phy.hr}
\affiliation{Department of Physics, Faculty of Science, 
	University of Zagreb, Bijeni\v{c}ka c. 32, 10000 Zagreb, Croatia}

\author{Bruno Klajn}
\affiliation{Department of Physics, Faculty of Science, 
	University of Zagreb, Bijeni\v{c}ka c. 32, 10000 Zagreb, Croatia}

\author{Dario Juki\'c}
\affiliation{Faculty of Civil Engineering, University of Zagreb, A. Ka\v{c}i\'{c}a  Mio\v{s}i\'{c}a 26, 10000 Zagreb, Croatia}

\author{Hrvoje Buljan}
\affiliation{Department of Physics, Faculty of Science, 
University of Zagreb, Bijeni\v{c}ka c. 32, 10000 Zagreb, Croatia}

\begin{abstract}
We study a system of strongly interacting one-dimensional (1D) bosons on a ring pierced by a synthetic magnetic flux tube. By the Fermi-Bose mapping, this system is related to the system of spin-polarized non-interacting electrons confined on a ring and pierced by a solenoid (magnetic flux tube). On the ring there is an external localized delta-function potential barrier $V(\phi)=g \delta(\phi-\phi_0)$. 
We study the Berry phase associated to the adiabatic motion of delta-function barrier around the ring as a function of the strength of the potential $g$ and the number of particles $N$. The behavior of the Berry phase can be explained via quantum mechanical reflection and 
tunneling through the moving barrier which pushes the particles around the ring. 
The barrier produces a cusp in the density to which one can associate a missing charge $\Delta q$ (missing density) for the case of electrons (bosons, respectively). We show that the Berry phase (i.e., the Aharonov-Bohm phase) cannot be identified with the quantity $\Delta q/\hbar \oint \mathbf{A}\cdot d\mathbf{l}$. This means that the missing charge cannot be identified as a (quasi)hole. We point out to the connection 
of this result and recent studies of synthetic anyons in noninteracting systems. In addition, for bosons we study the weakly-interacting regime, which is related to the strongly interacting electrons via Fermi-Bose duality in 1D systems. 
\\
\end{abstract}

\maketitle
\section{Introduction}

Exactly solvable one-dimensional (1D) quantum many-body models provide an insight into the 
strongly-correlated states not accessible with numerical simulations. 
The Lieb-Liniger (LL) model, which describes a 1D Bose gas with repulsive contact interactions of 
strength $c$, is solved with the Bethe ansatz~\cite{Lieb}. 
Experiments with ultracold atoms loaded in tight, transversely confined, 
effectively 1D atomic waveguides~\cite{experimental_1Dboson, Kinoshita_Science, Paredes, 
Kinoshita_Nature} have revived the interest in the LL model (see Ref.~\cite{Cazalilla2011} for a review). 
For infinite interaction strength ($c\rightarrow \infty$), the LL bosons enter the Tonks-Girardeau (TG) regime, 
where solutions are found by the Fermi-Bose mapping~\cite{Girardeau}. 
The TG regime has been experimentally achieved~\cite{Kinoshita_Science, Paredes, Kinoshita_Nature} with atoms 
at low temperatures and linear densities, and with strong effective interactions~\cite{Olshanii, Petrov, Dunjko}.

The developments of synthetic gauge fields for ultracold atoms have opened the way for 
investigating topological states of matter in these systems~\cite{Abo-Shaeer2001, Schweikhard2004, Struck2012, 
Miyake2013, Aidelsburger2013, Jotzu2014, Kennedy2015, Dalibard2011, Goldman2014, Cooper2019}. 
The single-particle topological phenomena are well understood~\cite{Goldman2014, Cooper2019}. 
However, strongly interacting quantum systems coupled to gauge fields can yield intriguing 
correlated topological states of matter, which are difficult to understand. 
It is natural to ask whether exactly solvable models coupled to gauge fields can provide some insight. 
We are interested in 1D quantum particles on a ring, which is pierced with a synthetic 
magnetic flux-tube (in this geometry the pertinent gauge field cannot be gauged out), and explore the 
Berry phase~\cite{Berry} as the quantum gas is stirred around the ring with an external local potential.  

This geometry is readily found in atomtronics - emerging field in quantum technology seeking for ultracold-gas analogs of electronic devices and circuits~\cite{Seaman}. An important example of an atomtronic circuit is provided by a Bose-Einstein condensate flowing in a ring-shaped trapping potential, which can be realized using different methods~\cite{Ryu2007, Ramanathan, Moulder, Marti, Henderson, Garraway}. Such systems interrupted by one or several weak barriers and pierced by an effective magnetic flux, have been studied in analogy with the superconducting quantum interference devices (SQUIDs)~\cite{Wright, Ramanathan, Ryu2013, Eckel, Haug_rPRA, Haug_QSci, Yakimenko, Aghamalyan2015, Aghamalyan2016, Schenke}. 
In particular, in system with weak barriers and weak atom-atom interaction, hysteresis effects have been evidenced~\cite{Eckel}.
The persistent current phenomenon has been theoretically characterized for 1D bosons in this geometry, for all interaction and barrier strengths~\cite{Cominotti}. 
Studies of the Aharonov-Bohm (AB) effect~\cite{Aharonov-Bohm} for the density excitations propagated through the ring predicted the absence of the AB oscillations for all interaction regimes~\cite{Tokuno2008, Haug_rPRA, Haug_QSci}. 
The presence of disorder leads to crossover from AB to Al'tshuler-Aronov-Spivak oscillations, investigated in the presence of bosonic interaction~\cite{Chretien}. 
This configuration can also serve to study the dynamics of vortices in a quantum fluid~\cite{Yakimenko}.
%%%%%%%%%%%%%%%%%%%%%%
For stronger interactions and higher barriers, Bose gas confined to a ring shaped lattice, has shown the emergence of the effective two-level system of current states, suggesting it to be a cold-atom analog of qubit~\cite{Aghamalyan2015,Aghamalyan2016}. Moreover, the study of bosonic Josephson effect in this geometry, has shown that strongly correlated 1D bosonic system exhibits the damping of the particle-current oscillations~\cite{Polo,Schenke}.
\\
Here we study the Berry phase in a system of $N$ strongly interacting 1D bosons on a ring of length 
$2\pi R$ subjected to the synthetic vector potential of a thin solenoid piercing the ring. 
Using the Fermi-Bose mapping, this system is related to the spin-polarized non-interacting 
electrons confined on a ring, pierced by a thin solenoid.
On the ring there is a localized delta-function barrier $V(\phi)=g\delta(\phi-\phi_0)$, $-\pi \leq \phi,\phi_0 \leq\pi$. 
We study the Berry phase when this external potential is adiabatically moved around the ring. 
First, we look at one particle in this configuration and find analytically the eigenstates of the Hamiltonian. 
We calculate the acquired Berry phase as a function of the strength of the potential $g$. 
Next, we consider a system of $N$ strongly interacting (impenetrable) bosons and $N$ noninteracting electrons. 
We calculate the Berry phase in dependence on the strength of the potential $g$ and the number of particles $N$. 
We also study bosons in the weakly interacting regime by using the Gross-Pitaevskii theory. 
The behavior of the Berry phase can be explained via quantum mechanical reflection and tunneling 
of the particles through the barrier, as it pushes them around the ring.

The delta-function barrier induces a cusp in the density to which one can relate a missing density $\Delta q$ (missing charge) for the case of bosons (electrons, respectively).
We show that the Berry phase cannot be identified with the quantity $\Delta q/\hbar \oint \textbf{A}\cdot d\textbf{l}$, and conclude that the missing density (charge) cannot be identified as a (quasi)hole. 
This exact result provides insight into a recent study of synthetic anyons in noninteracting systems~\cite{Lunic}. 
More specifically, when fractional flux-tubes pierce two-dimensional electron gas in the 
integer quantum Hall (IQH) state, braiding properties of these flux tubes are equivalent to those of anyons~\cite{Lunic, Weeks2007}. 
However, local perturbations in the density around the flux tubes cannot be identified as emergent quasiparticles~\cite{Lunic},
which is corroborated by this study in 1D quantum systems. 

%%%%%%%%%%%%%%%%%%%%%%%%%%%%%%%%%%%%%
%%%%%%%%%%%%%%%%%%%%%%%%%%%%%%%%%%%%%
\section{Berry phase for one particle on a ring}

We start by considering a particle confined on a ring of radius $R$, containing a localized delta-function 
potential barrier somewhere on the ring. 
This particle can be a boson of a synthetic charge $q$ subjected to a synthetic gauge field of a solenoid carrying flux 
$\Phi$ placed in the center of a ring, or an electron of electric charge $q$ coupled with a vector potential of a solenoid with a 
magnetic flux $\Phi$. In the rest of the paper we will refer to $q$ and $\Phi$ as to charge and flux, and we will not distinguish the electric (i.e., real) from the artificial charge and flux which can be engineered in ultracold atomic gases. This system is described by the Hamiltonian
\begin{equation}
 H=\frac{1}{2m}
 \left( -\frac{i\hbar}{R}\frac{\partial}{\partial \phi} - \frac{q\Phi}{2\pi R}\right)^2 + \bar{g}\delta(\phi-\phi_0),
\end{equation}
where $\phi\in [-\pi, \pi]$.
We introduce the dimensionless parameters $\alpha=q\Phi/h$ and $g=(2mR^2/\hbar^2)\bar{g}>0$, and dimensionless energy $\epsilon=(2mR^2/\hbar^2) E$. Our task is to solve time-independent dimensionless Schr\"odinger equation
\begin{equation}
-\left[ \left( \frac{\partial}{\partial \phi}-i\alpha\right)^2-g\delta(\phi-\phi_0)\right]\psi=\epsilon \psi.
\label{eq:dimensionless_sch_eq}
\end{equation}
For $\phi\neq\phi_0$, the delta term vanishes. Thus, for $\phi\in[-\pi, \phi_0\rangle$ we have
\begin{equation*}
\psi_l=Ae^{+i(\sqrt{\epsilon}+\alpha)\phi}+Be^{-i(\sqrt{\epsilon}+\alpha)\phi},
\end{equation*}
and for $\phi\in\langle\phi_0,\pi]$,
\begin{equation*}
\psi_r=Ce^{+i(\sqrt{\epsilon}+\alpha)\phi}+De^{-i(\sqrt{\epsilon}-\alpha)\phi}.
\end{equation*}
For the whole domain we write
\begin{equation}
    \psi=\theta(\phi_0-\phi)\psi_l+\theta(\phi-\phi_0)\psi_r.
    \label{eq:single_particle_function}
\end{equation}
Next, we impose boundary conditions: 
continuity of the wave function $\psi_l(-\pi)=\psi_r(+\pi)$, continuity of its derivative $\psi_l'(-\pi)=\psi_r'(+\pi)$, 
and continuity of the wave function at $\phi_0$. 
This leads us to the result
\begin{equation*}
\begin{split}
    \psi_l=&\mathcal{N}e^{i\alpha(\phi+\pi)}\big(e^{+i\sqrt{\epsilon}(\phi+\pi-\phi_0)}\sin[\pi(\sqrt{\epsilon}-\alpha)]\\
    &+e^{-i\sqrt{\epsilon}(\phi+\pi-\phi_0)}\sin[\pi(\sqrt{\epsilon}+\alpha)]\big),
\end{split}
\end{equation*}
and
\begin{equation*}
\begin{split}
     \psi_r=&\mathcal{N}e^{i\alpha(\phi-\pi)}\big(e^{+i\sqrt{\epsilon}(\phi-\pi-\phi_0)}\sin[\pi(\sqrt{\epsilon}-\alpha)]\\
     &+e^{-i\sqrt{\epsilon}(\phi-\pi-\phi_0)}\sin[\pi(\sqrt{\epsilon}+\alpha)]\big),
\end{split}
\end{equation*}
where $\mathcal{N}$ is the normalization constant:
\begin{equation}
\begin{split}
    \mathcal{N}=&\big [2\pi\big(1-\cos 2\pi\alpha\cos 2\pi\sqrt{\epsilon}\\
    &+\frac{\sin{2\pi\sqrt{\epsilon}}}{2\pi\sqrt{\epsilon}} (\cos 2\pi\alpha-\cos 2\pi\sqrt{\epsilon}\big)\big) \big]^{-1/2}.
    \label{eq:normalization}
\end{split}
\end{equation}
The energy $\epsilon$ can be found by integrating Eq. (\ref{eq:dimensionless_sch_eq}) around $\phi_0$, 
which yields $\psi_r'(\phi_0)-\psi_l'(\phi_0)=g\psi(\phi_0)$, i.e., an implicit equation for the energy:
\begin{equation}
    \cos 2\pi\alpha - \cos 2\pi \sqrt{\epsilon}=g\frac{\sin 2\pi\sqrt{\epsilon}}{2\sqrt{\epsilon}}.
    \label{eq:energies}
\end{equation}
Note that for $\alpha=0,1,2,\ldots$ the energy spectrum is mapped onto itself, which means that these cases are 
related by a simple gauge. Therefore it is sufficient to consider flux in the domain $\alpha \in [0,1]$. 

We are interested in the Berry phase~\cite{Berry} $\gamma$ when the delta-function travels adiabatically around the ring: 
$\phi_0\rightarrow \phi_0+2\pi$.
The Berry phase is 
\begin{equation}
    \gamma=\Delta +i\int_{-\pi}^{+\pi} d\phi_0\int_{-\pi}^{+\pi}d\phi \psi^*\frac{\partial}{\partial\phi_0}\psi,
    \label{eq:Berry_phase}
\end{equation}
where $\Delta$ denotes the phase difference of the wave function when parameter $\phi_0$ is at the endpoints of a closed path~\cite{Mukunda,Berry}. Namely, the wave function is a single-valued function of the variable $\phi$, but multivalued in the parameter $\phi_0$. 
The phases of the wave function at endpoints $\pm \pi$ differ as 
\begin{equation*}
    \frac{\psi(\phi_0=+\pi)}{\psi(\phi_0=-\pi)}=e^{2\pi i \alpha},
\end{equation*}
i.e. $\Delta=2\pi\alpha$. By calculating the derivatives, we obtain 
\begin{equation}
    \gamma = 2\pi\alpha-(2\pi\mathcal{N})^2\sqrt{\epsilon}\sin{2\pi\alpha}\sin{2\pi\sqrt{\epsilon}}.
    \label{eq:Berry_phase_calculated}
\end{equation}

%%%%%%%%%%%%%% FIGURE %%%%%%%%%%%%%%%%%%%
\begin{figure}
    \centering
    \includegraphics[width=0.47\textwidth]{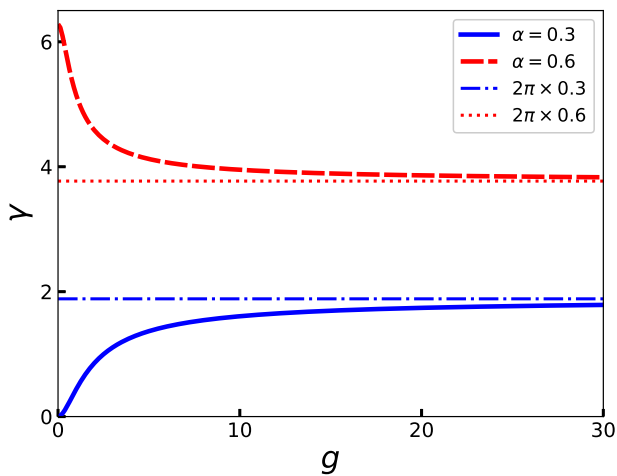}
    \caption{The Berry phase as a function of the barrier height $g$ for one particle on a ring.
    The parameter $\alpha$ describing the flux through the ring is $\alpha=0.3$ and $\alpha=0.6$.
    Horizontal lines denote the asymptotic value of the Berry phase $q\Phi/\hbar = 2\pi \alpha$ for $g\rightarrow \infty$.}
    \label{fig:berry_phase_particle_on_a_ring}
\end{figure}
%%%%%%%%%%%%%%%%%%%%%%%%%%%%%%%%%%%%%%%%%
The dependence of the Berry phase on the height of the potential barrier is shown in Fig. \ref{fig:berry_phase_particle_on_a_ring}.
For the vanishing barrier, Eq. (\ref{eq:energies}) and Eq. (\ref{eq:Berry_phase_calculated}) give $\gamma=0$ when $\alpha<0.5$, and $\gamma=2\pi$ when $\alpha>0.5$. Both results describe vanishing Berry phase, as expected.
As the potential barrier becomes stronger, the Berry phase increases (decreases) for $\alpha<0.5$ ($\alpha>0.5$, respectively). 
In the limit of infinitely strong potential barrier $g\rightarrow \infty$, the Berry phase saturates at the value 
$\gamma=2\pi\alpha =q\Phi/\hbar$.
This result is equal to the Aharonov-Bohm (AB) phase~\cite{Aharonov-Bohm} acquired when {\em one} particle of charge 
$q$ circles around the solenoid carrying flux $\Phi$. 

Results presented in Fig. \ref{fig:berry_phase_particle_on_a_ring} can be explained through the phenomena of 
quantum-mechanical tunneling and reflection.
As the barrier moves, in the classical sense it pushes the particle; the particle can tunnel through, or be reflected from the barrier. 
Thus, the whole probability density (i.e., the whole charge of the particle) will generally not make a full circle 
around the ring, but only a part of it. 
The Berry phase is the AB phase acquired by the amount of probability density that encircled the flux tube. 
The particle probability density reflected from the moving barrier, also moves around the flux tube and acquires the AB phase. 
In contrast, the probability density that tunneled through the barrier does not contribute to the AB phase. 
For the infinite barrier there is total reflection, i.e., one particle of charge $q$ moved around the flux $\Phi$, 
resulting in the phase $q\Phi/\hbar$. 

Finally, we generalize our result and consider a situation where the solenoid of flux $\Phi$ is inside the ring, but at the distance $r<R$ from the center of the ring. 
It can be shown that the wave function $\psi_R$ for a displaced solenoid is related to the 
wave function $\psi$ by a gauge transformation,
\begin{equation*}
\psi_R=\psi\exp\big\{i\alpha\big[\arctan\bigg(\frac{R+r}{R-r}\tan{\frac{\phi}{2}\bigg)-\frac{\phi}{2}}\big]\big\}.
\label{eq:gauge_transformation}
\end{equation*}
Energy remains the same as in Eq. (\ref{eq:energies}); the Berry phase in Eq. (\ref{eq:Berry_phase_calculated}) is also unchanged since the additional gauge factor does not depend on $\phi_0$. Thus, our previous analysis is generally valid for a particle on ring threaded by a flux tube anywhere inside the ring. 

%%%%%%%%%%%%%%%%%%%%%%%%%%%%%%%%%%%%%
%%%%%%%%%%%%%%%%%%%%%%%%%%%%%%%%%%%%%
\section{Berry phase for strongly interacting bosons on a ring}

Now we consider a system of $N$ indistinguishable bosons interacting via point-like interactions in the same configuration, described by the Lieb-Liniger model~\cite{Lieb} with an additional gauge term:
\begin{equation}
\begin{split}
    H=&\sum_{i=1}^{N}\left[ \frac{1}{2m}
    \left( -\frac{i\hbar}{R}\frac{\partial}{\partial \phi_i} - \frac{q\Phi}{2\pi R}\right)^2 + \bar{g}\delta(\phi_i-\phi_0) \right]\\
    &+c_{1D}\sum_{1\leq i<j\leq N} \delta(\phi_i-\phi_j).
    \label{eq:general_Hamiltonian}
\end{split}
\end{equation}
Here $c_{1D}$ is the effective 1D interaction strength. By varying $c_{1D}$, the system can be tuned from the weakly interacting regime described by the mean field theory, up to the strongly interacting TG regime with infinitely repulsive contact interactions $c_{1D}\rightarrow \infty$. In the TG limit, the interaction term of the Hamiltonian can be replaced by a boundary condition on the many-body wave function~\cite{Girardeau, Cazalilla2011}

\begin{equation*}
    \Psi_{TG}(\phi_1,\phi_2,...,\phi_N,t)=0\quad \text{if}\quad \phi_i=\phi_j
\end{equation*}
for any $i\neq j$. Now, the Hamiltonian becomes 
\begin{equation*}
    H=\sum_{i=1}^{N}\left[ \frac{1}{2m}
    \left( -\frac{i\hbar}{R}\frac{\partial}{\partial \phi_i} - \frac{q\Phi}{2\pi R}\right)^2 + \bar{g}\delta(\phi_i-\phi_0) \right].
\end{equation*}
The bosonic many-body wave function $\Psi_{TG}$ satisfying the boundary condition and the Schr\"odinger equation is related to the fermionic wave function $\Psi_{F}$, which describes a system of $N$ noninteracting spinless fermions through the Fermi-Bose mapping~\cite{Girardeau}:
\begin{equation}
    \Psi_{TG}=\prod_{1\leq j<l\leq N} \text{sgn}(\phi_j-\phi_l)  \Psi_{F}.
    \label{eq:many_body_wf}
\end{equation}
Here, $ \Psi_{F}$ is given by the Slater determinant, 
\begin{equation*}
    \Psi_{F}=\frac{1}{\sqrt{N!}}\text{det}[\psi_k(\phi_j,t)],
\end{equation*}
where $\psi_k(\phi,t)$ denote $N$ orthonormal single-particle wave functions obeying a set of uncoupled single-particle Schr\"odinger equations
\begin{equation}
   i\hbar \frac{\partial \psi_k}{\partial t}=\left[\frac{1}{2m} \left( -\frac{i\hbar}{R}\frac{\partial}{\partial \phi} - \frac{q\Phi}{2\pi R}\right)^2 + \bar{g}\delta(\phi-\phi_0)\right] \psi_k.
    \label{eq:single_particle}
\end{equation}
The eigenfunctions of the single-particle Schr\"odinger equation are given in Eq. (\ref{eq:single_particle_function}) with normalization constant in Eq. (\ref{eq:normalization}). The energies of the single-particle states are given by Eq. (\ref{eq:energies}); bosons in the TG gas occupy states from the lowest energy state up to the $N$-th energy state. 

%%%%%%%%%%%%%% FIGURE %%%%%%%%%%%%%%%%%%%%%%%%%%
\begin{figure}
\includegraphics[width=0.47\textwidth]{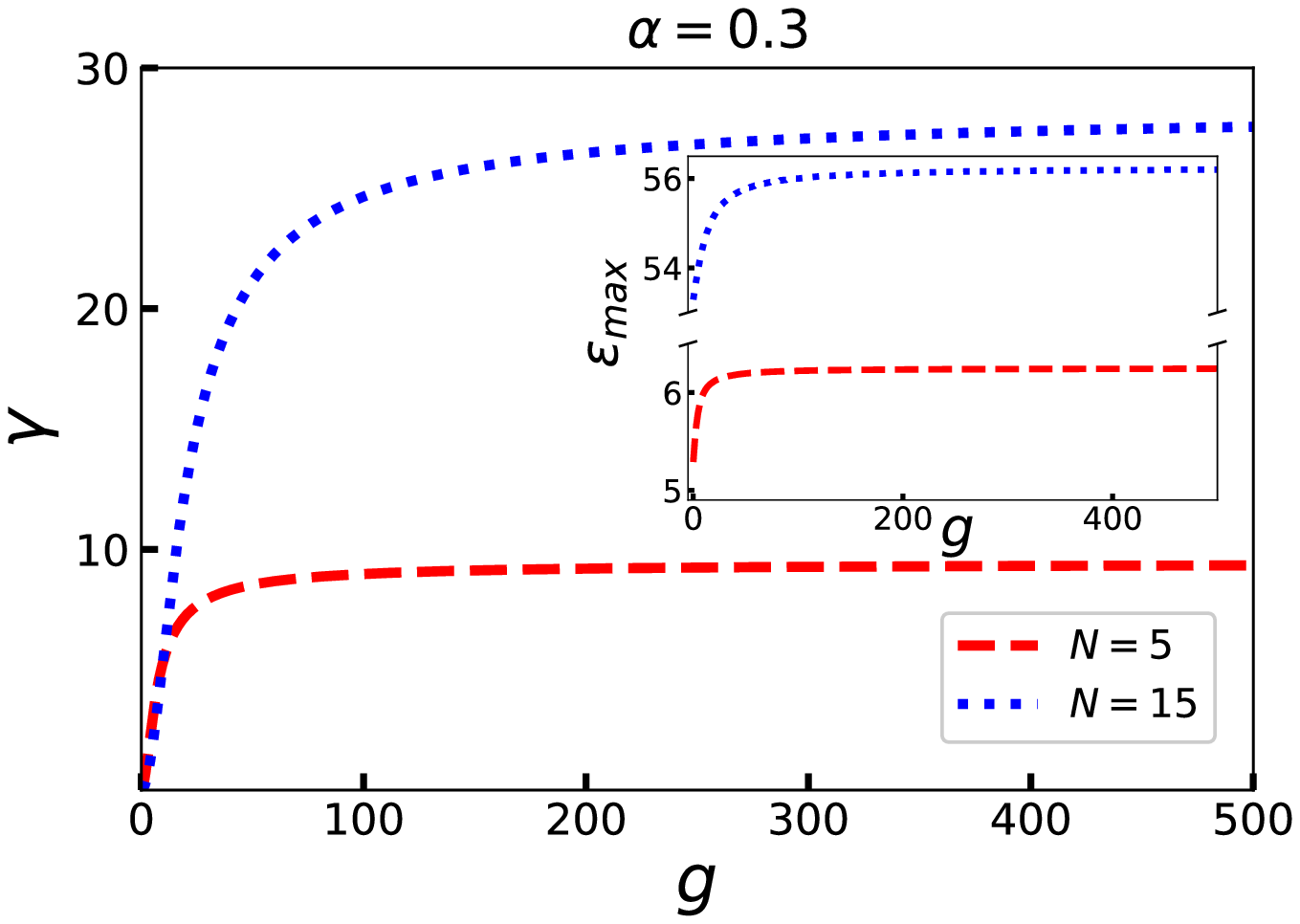}\\
\vspace{0.5cm}
\includegraphics[width=0.47\textwidth]{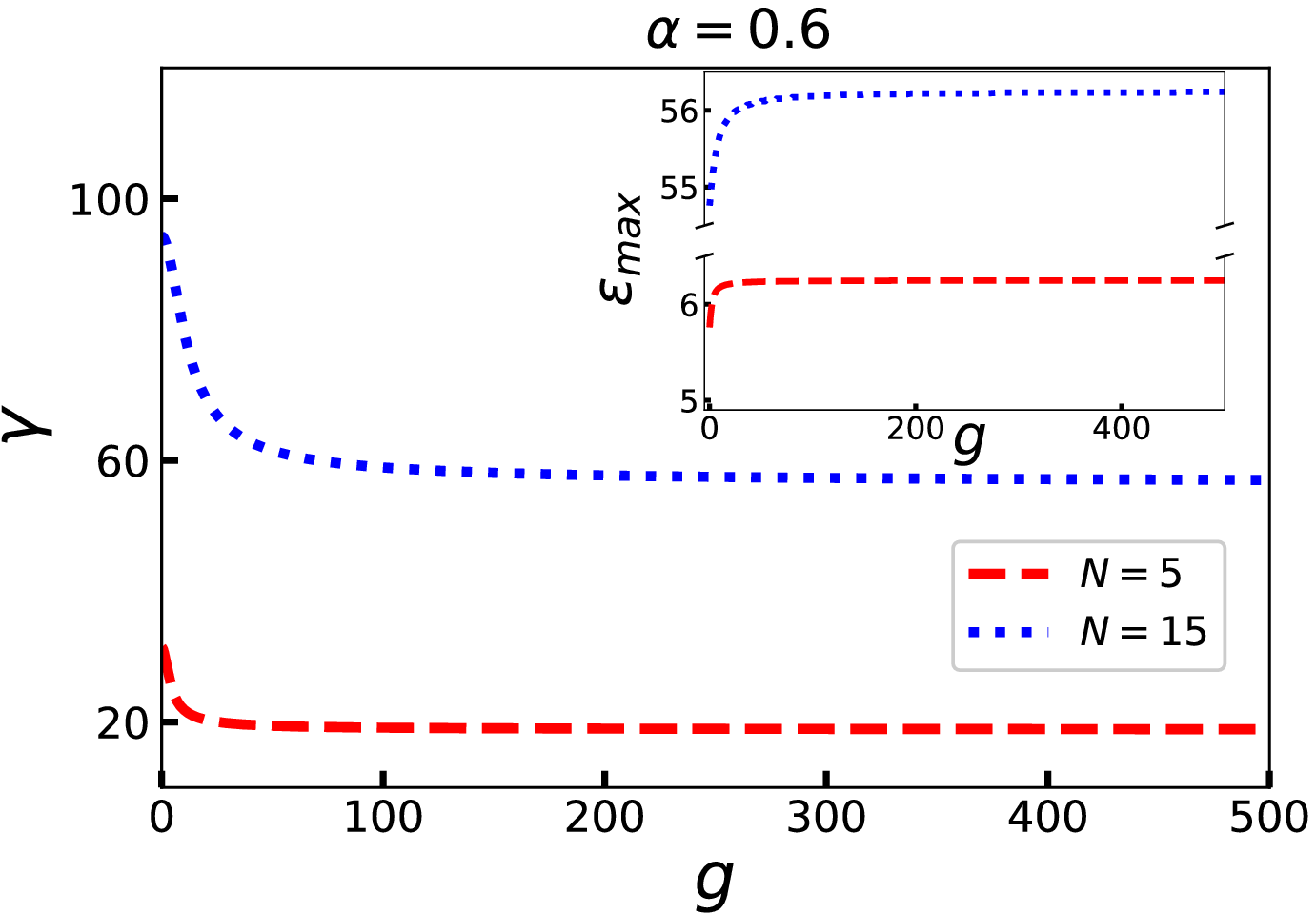}
\caption{Berry phase for $N$ Tonks-Girardeau bosons as a function of the strength of the barrier $g$ for $\alpha=0.3$ (upper plot) and $\alpha=0.6$ (lower plot). The insets show the energy of the highest occupied single-particle state as a function of $g$. }
\label{fig:Berry_V0_TG}
\end{figure}

We study now the Berry phase arising when the barrier potential is set into adiabatic anticlockwise rotation around the ring.
The Berry phase is
\begin{equation}
    \gamma=\Delta +i\int_{-\pi}^{+\pi} d\phi_0 \langle \Psi_{TG} | \frac{\partial}{\partial\phi_0} | \Psi_{TG} \rangle,
    \label{eq:Berry_phase}
\end{equation}
where $\Delta$ is the phase difference of the wave function at the endpoints~\cite{Mukunda,Berry}, which is for $N$ particles given by 
\begin{equation*}
    \Delta= N 2\pi\alpha.
\end{equation*}

The second term in Eq. (\ref{eq:Berry_phase}) is calculated by using the fact that $<\Psi_{TG}|\partial/\partial \phi_0|\Psi_{TG}>=<\Psi_{F}|\partial/\partial \phi_0|\Psi_{F}>$, i.e., one has to calculate the Berry phase for the Slater determinant wave function. This problem was studied in detail in~\cite{Resta1994, Resta2000}, 
where it was shown that the Berry phase is a sum over the Berry phases of single-particle states
\begin{equation*}
    \gamma=\Delta + i\sum_{n=1}^{N}\int_{-\pi}^{+\pi} d\phi_0  \langle \psi_n | \frac{\partial}{\partial \phi_0} | \psi_n \rangle.
\end{equation*}
In the previous section, we have already solved the one particle case in Eq. $(\ref{eq:Berry_phase_calculated})$, which leads to
\begin{equation}
    \gamma = N2\pi\alpha-\sum_{n=1}^{N}(2\pi\mathcal{N}_n)^2\sqrt{\epsilon_n}\sin{2\pi\alpha}\sin{2\pi\sqrt{\epsilon_n}}.
    \label{eq:Berry_phase_calculated_TG}
\end{equation}

The dependence of the Berry phase (\ref{eq:Berry_phase_calculated_TG}) on the strength of the potential barrier $g$ is illustrated in Fig. \ref{fig:Berry_V0_TG}, for different $N$ and $\alpha$.
We do not plot the phase modulo $2\pi$ for clearer view. 
For $g=0$, the Berry phase is zero or an integer of $2\pi$. By increasing the barrier strength, the Berry phase 
monotonically increases for $\alpha=0.3$ (decreases for $\alpha=0.6$), and saturates at the value 
\begin{equation}
    \gamma=N2\pi\alpha=N q\Phi/\hbar
    \label{eq:Berry_AB}
\end{equation}
in the limit $g\rightarrow \infty$. 
This is the AB phase collected when $N$ particles of charge $q$ circle around the solenoid with flux $\Phi$. Results in Fig. \ref{fig:Berry_V0_TG} can again be interpreted through the tunneling and reflection of the particle density from the moving barrier, in the same fashion as for a single-particle. 

Here we take into account that single-particle states that contribute to the Berry phase (\ref{eq:Berry_phase_calculated_TG}) have different energies, and consequently different transmission probabilities. In the inset of Fig. \ref{fig:Berry_V0_TG} we plot the highest single-particle energy contributing to the Berry phase. For large $g$ this energy saturates, confirming the behavior of the Berry phase on the plot.

%%%%%%%%%%%%%%%%%%%%%%%%%%%%%%%%%%%%%%%%%%%%%%%%%%%%%%%%%%%%
%%%%%%%%%%%%%%%%%%%%%%%%%%%%%%%%%%%%%%%%%%%%%%%%%%%%%%%%%%%%
\section{Missing density (missing charge) is not an emergent quasiparticle}

\begin{figure}
    \centering
    \includegraphics[width=0.47\textwidth]{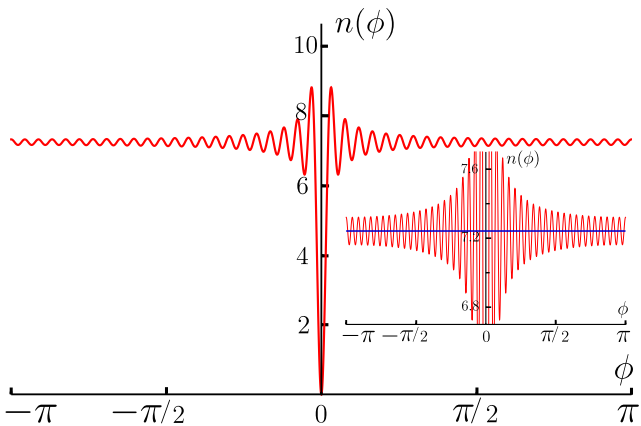}
    \caption{Single-particle density for $\alpha=0.3$, $N=45$ particles and infinite barrier $g=\infty$. Inset shows magnified view of single-particle density and horizontal blue line at $(N+1/2)/2\pi$.}
    \label{fig:single_particle_density}
\end{figure}

The single-particle density of TG gas described by Eq. (\ref{eq:many_body_wf}) is given as
$n(\phi)=\sum_{k=1}^N|\psi_k|^2$~\cite{Girardeau}. 
In Fig. \ref{fig:single_particle_density} we show the single-particle density when 
an impenetrable delta barrier is placed at $\phi_0=0$. 
At the position of the barrier, there is a cusp in the density.
For a sufficiently large number of particles, one can define a missing synthetic charge $\Delta q$ for the system of TG bosons, or the missing electric charge $\Delta q$ for noninteracting electrons on the Fermi side of the mapping.

%%%%%%%%%%%%%%%%%%%%%%%%%%%%%%%%%
%%%%%%%%%% FIGURE %%%%%%%%%%%%%%%%%%%
%%%%%%%%%%%%%%%%%%%%%%%%%%%%%%%%%

We calculate the missing charge in the thermodynamic limit, $N\rightarrow\infty$, $R\rightarrow \infty$, $N/2\pi R=\tilde n_{0}$. 
The coordinate space is now $x=R\phi\in\langle-\infty,\infty\rangle$. 
If there is no barrier, the particle density is uniform and equal to $\tilde n_{0}$. 
For simplicity, suppose that we insert an impenetrable barrier with $g\rightarrow \infty$ at $x=0$. 
In this limit, it is straightforward to calculate the single particle density,
\begin{equation}
\tilde n(x)=\tilde n_{0}-\frac{\sin{(2\pi \tilde n_{0} x)}}{2\pi x}.
\end{equation}
The missing charge is 
\begin{equation}
    \Delta q = q \int_{-\infty}^{\infty}[\tilde n(x)-\tilde n_{0}]dx=-\frac{q}{2}.
    \label{eq:missing_charge}
\end{equation}
Thus, the barrier induces density fluctuations which carry fractional charge $-q/2$.

In order to shed more light onto this result, we return to the geometry of the (finite) ring. 
For $N$ particles on the ring, in the absence of the barrier, the angular density is $n_0=N/2\pi$. 
We then insert an impenetrable barrier at $\phi_0=0$. 
The energy of the $k$-th single-particle state for $g=\infty$ is $\epsilon=k^2/4$, $k=1,2,\ldots$, 
and the single particle density is
\begin{eqnarray}
n(\phi) & = & \sum_{k=1}^N|\psi_k|^2= \sum_{k=1}^N \frac{1}{2\pi }(1-\cos k\phi) \nonumber\\
         & = & \frac{N+1/2}{2\pi}-\frac{1}{4\pi}\left( \frac{\sin (N\phi)}{\tan(\phi/2)}+\cos(N\phi)\right). 
\label{eq:linear_density}
\end{eqnarray}
The density $n(\phi)$ integrated over the ring gives $N$ particles, i.e., the number of particles 
on the ring is unchanged after we insert the delta barrier.
The first term in Eq. (\ref{eq:linear_density}), i.e., $(N+1/2)/2\pi$, corresponds to the uniform density of 
$N+1/2$ particles, and the second term gives density fluctuations of the missing $(-1/2)$ charge, 
in agreement with the fact that the number of particles does not change after insertion of the barrier. 
This is supported with the inset in Fig. \ref{fig:single_particle_density}, where we show the 
single-particle density $n(\phi)$, and the horizontal line at $(N+1/2)/2\pi$, which goes 
through the center of the density oscillations away from the barrier. 

Note that we cannot use a formula analogous to (\ref{eq:missing_charge}) to calculate 
the missing charge on the ring, simply because $\int_{-\pi}^{\pi}[n(\phi)-n_0]d\phi=0$, i.e., 
the number of particles does not change as we insert the delta barrier. 
One could try to resort to a formula such as $\int_{-\phi^*}^{\phi^*}[n(\phi)-n_0]d\phi=0$, 
i.e., to integrate over a region around the density dip induced by the barrier, but it 
is difficult to unambiguously define the region of integration $[-\phi^*,\phi^*]$ because 
the decay of the density oscillations is algebraic, i.e., without a scale. 
However, the thermodynamic limit allows for an unambiguous calculation of the missing charge 
via Eq. (\ref{eq:missing_charge}), because in this limit $(N+1/2)/2\pi R=N/2\pi R=\tilde n_{0}$. 

It may be tempting to interpret the obtained missing fractional charge as a 
fractional quasiparticle. 
When the delta barrier moves around the ring, one may consider the Berry phase 
(or Aharonov-Bohm phase for electrons) as the phase acquired by the motion of the missing charge. 
If the missing charge was caused by a quasiparticle excitation, this picture would be correct, 
however, this is not the case. The Berry phase acquired for a barrier with $g=\infty$ is $N q\Phi/\hbar$. 
On the other hand, the AB phase acquired by the motion of the missing charge around the 
ring is $\Delta q\Phi/\hbar$. Since $\Delta q \Phi/\hbar \neq Nq \Phi/\hbar$ modulo $2\pi$, 
we conclude that one cannot interpret the Berry phase as the motion of the missing charge, 
but rather as the movement of the particles reflected from the barrier as it pushes them around the solenoid. 
The cusp in the density cannot be considered as a quasiparticle. 

While this conclusion seems clear and perhaps obvious in this 1D system, we 
find that it provides insight into studies of braiding of fractional fluxes in 2D electron gases 
in magnetic fields~\cite{Lunic, Weeks2007}. More specifically, consider a 2D electron gas in a magnetic field 
in the IQH state. When this system is pierced with flux-tubes carrying fractional fluxes, 
it can be shown that braiding of fractional fluxes has anyonic properties~\cite{Lunic}. 
One can ask whether the missing charge around these fluxes behaves as a quasiparticle~\cite{Weeks2007} or not~\cite{Lunic}.
We have found, in consistency with this report, that the missing charge around these fluxes cannot be considered as a quasiparticle~\cite{Lunic}.

%%%%%%%%%%%%%%%%%%%%%% Gross-Pitaevskii %%%%%%%%%%%%%%%%%%
\section{Berry phase for weakly interacting bosons on a ring}

Now we turn to the weakly interacting regime described by the Gross-Pitaevskii theory (e.g., see Ref.~\cite{Dalfovo}). 
The GP equation for our problem is given by 
\begin{equation}
\begin{split}
    &\frac{1}{2m}\left(-\frac{i\hbar}{R}\frac{\partial}{\partial\phi}-\frac{q\Phi}{2\pi R}\right)^2\psi + \bar{g}\delta(\phi-\phi_0)\psi\\
    &+c_{1D}N| \psi |^2\psi =\mu \psi(\phi; \phi_0),
    \label{eq:GP_equation}
\end{split}
\end{equation}
where $\mu$ is the chemical potential. Without loss of generality, we assumed that the solenoid is placed in the center of the ring.
The effect of interactions is contained in a non-linear mean field term. We are interested in the behavior of the Berry phase in dependence of the strength of the mean field interaction $c_{1D}N$. 

We calculate the Berry phase numerically following Ref.~\cite{Mukunda}. 
The delta barrier is approximated as a rectangular potential barrier. The evolution parameter, angle $\phi_0$, is discretized to obtain a set of $T$ equidistant points denoted by $\phi_0(t)$. The wave function $\psi(\phi; \phi_0(t))$, corresponding to the barrier position at $\phi_0(t)$, 
is the lowest single-particle eigenstate found by diagonalization of Eq. (\ref{eq:GP_equation}). 
The overlap at two different points is $M(k,l)=\langle\psi(\phi; \phi_0(k))|\psi(\phi; \phi_0(l))\rangle$, and the product
\begin{equation*}
    U=M(0,1)M(1,2)...M(T,0)
\end{equation*}
gives the Berry phase
\begin{equation*}
    \gamma_0= -\text{arg}(U).
\end{equation*}
Note that $\gamma_0$ is the Berry phase per particle since we discuss now the mean field regime. 
The mean-field many-body wave function is given by $\prod_{i=1}^N\psi(\phi_i; \phi_0)$, from 
which we find the Berry phase $\gamma=N\gamma_0$.

%%%%%%%%%%%%%%%%%%%%%%%%%%%%%%%%%%%%
%%%%%%%%%%%%%%%%%%%%%%%%%%%%%%%%%%%%
\begin{figure}[!ht]
    \centering
    \includegraphics[width=0.47\textwidth]{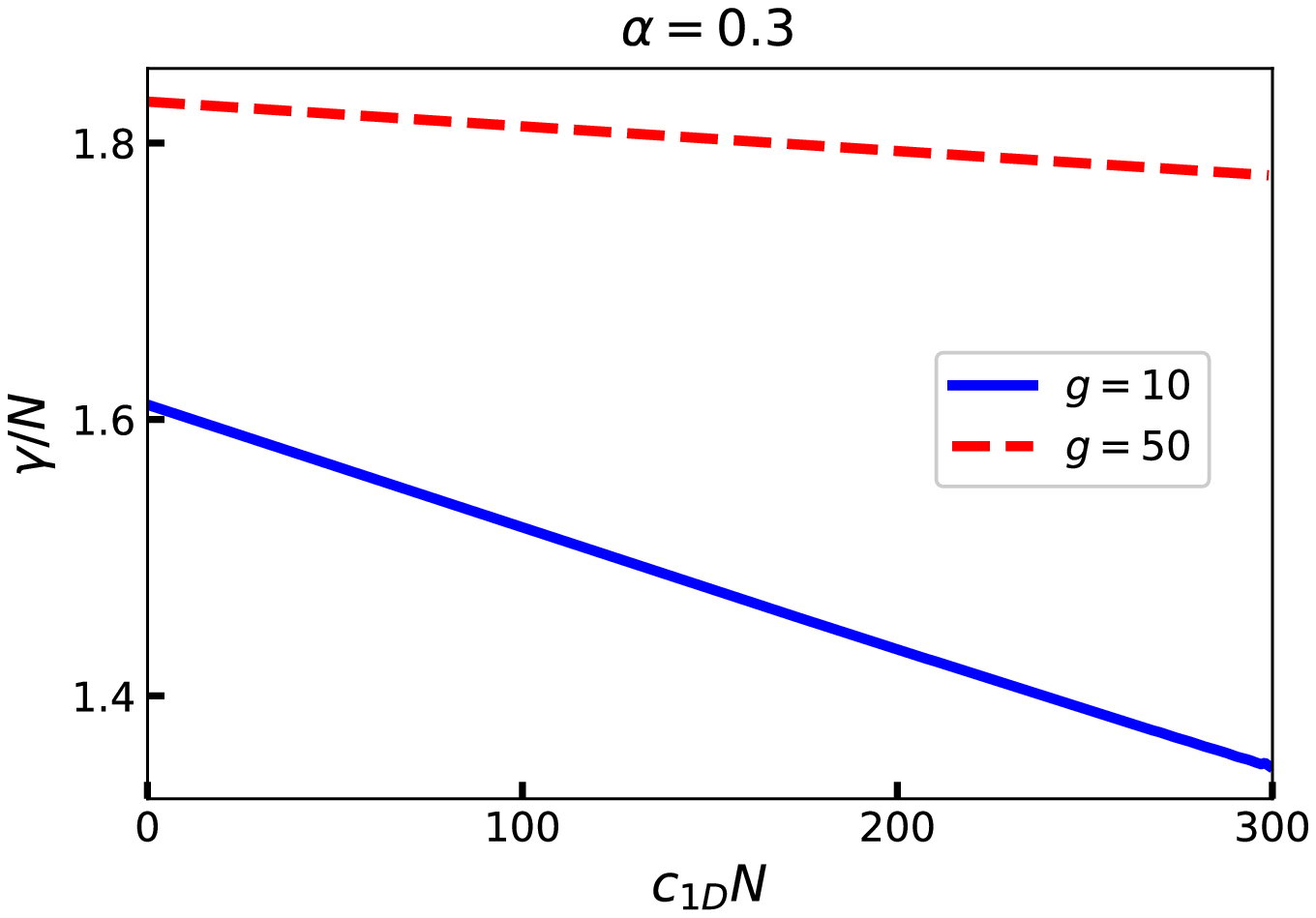}\\
    \vspace{0.5cm}
    \includegraphics[width=0.47\textwidth]{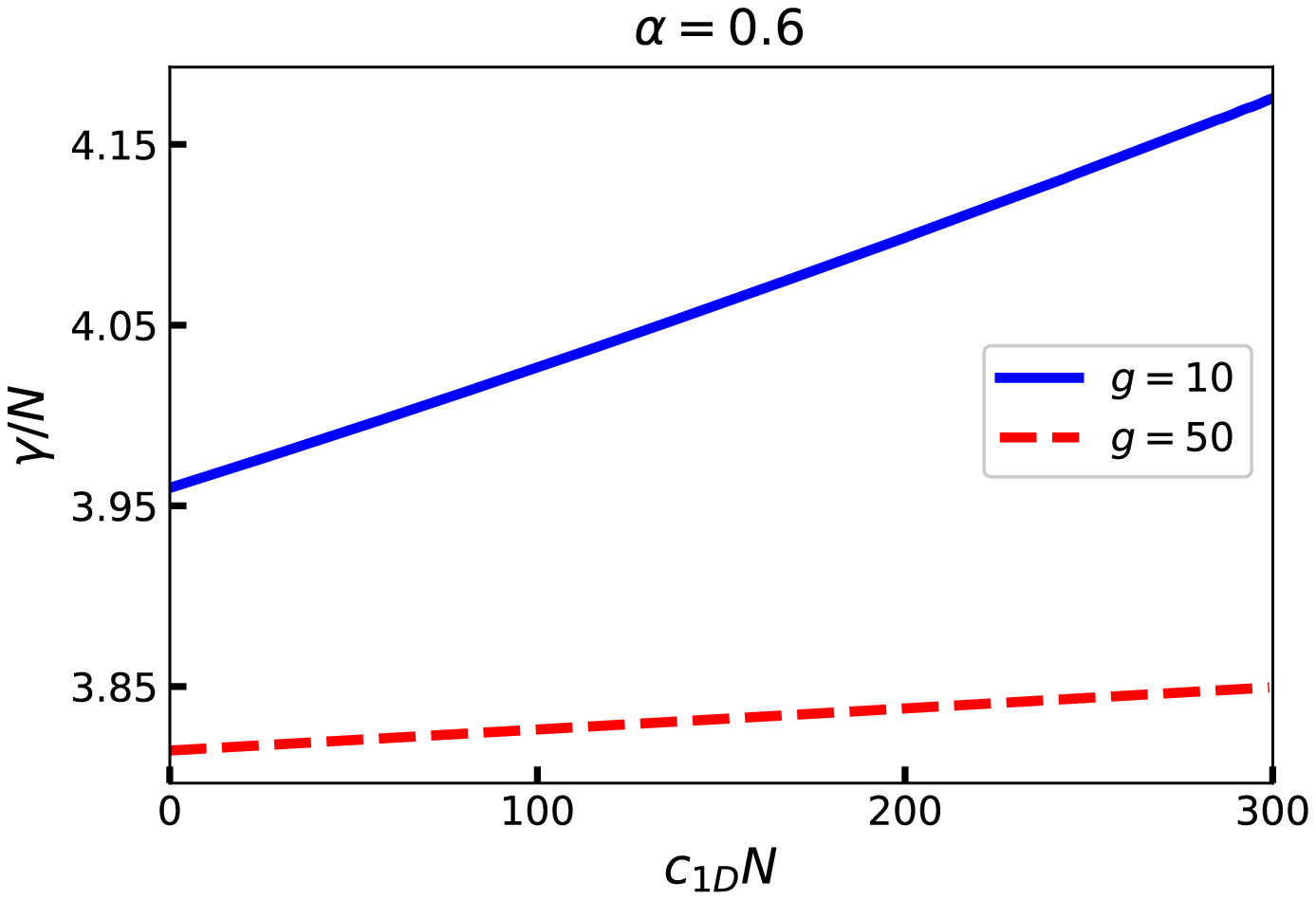}
    \caption{Berry phase per particle as a function of $c_{1D}N$ calculated in the mean field regime. 
    Results are shown for two values of the flux: $\alpha=0.3$ and $\alpha=0.6$, and two different barriers 
    corresponding to $g=10$ and $g=50$.}
    \label{fig:Berry_eff_int}
\end{figure}
%%%%%%%%%%%%%%%%%%%%%%%%%%%%%%%%%%%%
%%%%%%%%%%%%%%%%%%%%%%%%%%%%%%%%%%%%

In Fig. \ref{fig:Berry_eff_int} we show the dependence of the Berry phase per particle on 
the strength of the effective potential $c_{1D}N$, for different barrier strengths. 
With the increase of $c_{1D}N$, the chemical potential increases as well. 
This means that, effectively, increase of $c_{1D}N$ should lead to the same trend in 
the behavior of the Berry phase as the decrease of the barrier strength $g$, since the states with higher energy tunnel more easily through the barrier. We see that this is indeed the case by comparing Fig. \ref{fig:berry_phase_particle_on_a_ring} and Fig. \ref{fig:Berry_eff_int}. For $\alpha=0.3$ ($\alpha<0.5$), the Berry phase per particle $\gamma/N$ decreases with the increase of $c_{1D}N$; the same trend occurs when $g$ is decreased for a single particle at $\alpha=0.3$ as 
depicted in Fig. \ref{fig:berry_phase_particle_on_a_ring}. 
For $\alpha=0.6$ ($\alpha>0.5$), we see that $\gamma/N$ increases with the increase of $c_{1D}N$; 
the same trend occurs when $g$ is decreased for a single particle at $\alpha=0.6$.
%see Fig. \ref{fig:berry_phase_particle_on_a_ring}. 
This is consistent with the interpretation of the Berry phase via reflection and transmission 
of the particles through the moving barrier. 

%%%%%%%%%%%%%%%%%%%%%%%%%%%%%%%%%%%%%
%%%%%%%%%%%%%%%%%%%%%%%%%%%%%%%%%%%%%
\section{Conclusion}

In conclusion, we have studied the Berry phase in a system of interacting 1D bosons on a ring, 
with an external localized delta-function potential on the ring, and a synthetic solenoid threading the ring. 
We have calculated the Berry phase associated to the adiabatic motion of the delta-function potential 
around the ring. Results are shown for a single particle, for the impenetrable Tonks-Girardeau bosons
(where identical results hold for noninteracting spinless electrons via Fermi-Bose mapping), 
and interacting bosons in the Gross-Pitaevskii mean field regime. 
The behavior of the Berry phase can be explained via quantum mechanical reflection and 
tunneling through the moving barrier which pushes the particles around the ring. 
For an impenetrable barrier, the Berry phase is given by $Nq\Phi/\hbar$, where $q$ is the 
synthetic charge of one particle, $\Phi$ is the flux through the solenoid, and $N$ is the 
number of particles. These results provide insight into systems of BECs in toroidal traps 
used in the context of atomtronics. 

In addition, our results provide insight into the interpretation of the Berry phase 
obtained when fractional fluxes piercing a 2D electron gas in the IQH state are braided~\cite{Lunic}. 
An infinite barrier expels the particle density away from itself, 
leading to a cusp in the density profile, to which one can associate a missing density, i.e., 
a missing charge $\Delta q$.
We have shown that the Berry phase cannot be identified with the quantity 
$\Delta q/\hbar \oint \mathbf{A}\cdot d\mathbf{l}$, which shows that 
the missing density (charge) cannot be identified as a (quasi)hole. 

\section{Acknowledgments}
We are grateful to Anna Minguzzi for useful comments. This work was supported by the Croatian Science Foundation Grant No. IP-2016-06-5885 SynthMagIA, and in part by the QuantiXLie Center of Excellence, a project co-financed by the Croatian Government and European Union through the European Regional Development Fund - the Competitiveness and Cohesion Operational Programme (Grant KK.01.1.1.01.0004).

\end{document}